\documentclass[%
 aip,
 amsmath,amssymb,
 reprint,%
]{revtex4-1}

\usepackage{graphicx}
\usepackage{dcolumn}
\usepackage{bm}

\usepackage[utf8]{inputenc}
\usepackage[T1]{fontenc}
\usepackage{mathptmx}
\usepackage{etoolbox}

\makeatletter
\def\@email#1#2{%
 \endgroup
 \patchcmd{\titleblock@produce}
  {\frontmatter@RRAPformat}
  {\frontmatter@RRAPformat{\produce@RRAP{*#1\href{mailto:#2}{#2}}}\frontmatter@RRAPformat}
  {}{}
}%
\makeatother
\begin{document}

\preprint{AIP/123-QED}

\title[]{Diffusion-entropy scaling across dimensions}
\author{Nayana Venkatareddy}
 \affiliation{Department of Physics, Indian Institute of Science, C. V. Raman Ave, Bengaluru 560012, India}
 \author{Mohd Moid}%
\affiliation{Department of Fundamental Engineering, Institute of Industrial Science, University of Tokyo, Tokyo, Japan}
\author{Prabal K. Maiti }
\email{maiti@iisc.ac.in}
\affiliation{Department of Physics, Indian Institute of Science, C. V. Raman Ave, Bengaluru 560012, India}

\author{Biman Bagchi}
\email{bbagchi@iisc.ac.in}
\affiliation{Solid State and Structural Chemistry Unit, Indian Institute of Science, C. V. Raman Ave, Bengaluru 560012, India}

\date{\today}

\begin{abstract}

 A quantitative relationship between the diffusion coefficient \(D\) of a tagged particle in a liquid and the entropy \(S\) of that liquid has long been sought, as it would allow entropy to be inferred directly from diffusion measurements and transport properties to be predicted from thermodynamic information. Here, we employ extensive computer simulations to independently compute both \(D\) and \(S\) for Lennard-Jones (LJ) liquids and for water across a wide range of thermodynamic state points. Our study covers two and three dimensions for both systems, and additionally explores one-dimensional confinement for water.
We find that \textit{the ratio of diffusion coefficients between two states follows an almost perfect exponential dependence on their entropy difference}. For LJ liquids, the exponential prefactor exhibits a pronounced dependence on dimensionality \(d\), consistent in trend but quantitatively distinct from theoretical predictions. In contrast, \textit{water shows a strikingly weak dimensionality (d) dependence}, deviating from theory, which we attribute to the dominant role of jump diffusion.
Remarkably, the exponential diffusion–entropy relationship persists even when translational and rotational contributions to entropy are separated and considered individually. This robustness suggests that entropy provides a unifying measure governing particle mobility in liquids, largely independent of microscopic mechanisms or dimensional constraints.

\end{abstract}

\maketitle

\section{\label{sec:level1}Introduction}

Einstein’s 1905 work \cite{einstein1905motion} on Brownian motion laid the groundwork for viewing diffusion as a stochastic process driven by thermal fluctuations. This seminal contribution not only provided a microscopic foundation for Fick’s law but also established diffusion as a bridge between statistical mechanics and observable macroscopic transport. It helped frame the broader connection between thermodynamics and kinetics, a longstanding pursuit originating from the formulation of thermodynamic transition state theory \cite{10.1063/1.1749604}, which relates the rate of a reaction to the height of free energy barriers. Building on these ideas, Smoluchowski \cite{Smoluchowski+1918+129+168}, Kramers \cite{KRAMERS1940284}, and Eyring \cite{Glasstone1941TheTO} pioneered theoretical descriptions of barrier crossing in condensed phases, in which viscosity and diffusivity explicitly enter as determinants of kinetic rates. In these frameworks, the rate of relaxation or reaction is predicted to scale with the diffusion coefficient, thereby tying dynamical processes to transport properties.

Despite these advances, a direct and quantitative relation between diffusion and entropy remains far from obvious. Entropy, as a thermodynamic state function, quantifies the accessible phase space of the system, while diffusion reflects the dynamical rate of exploring this phase space. Over the years, several studies have attempted to bridge this conceptual divide \cite{YaakovRosenfeld_1999,Dzugutov1996}. A landmark development came from Kauzmann’s \cite{Kauzmann1948} discussion of the “entropy crisis” in supercooled liquids, which highlighted the central role of entropy in controlling relaxation and glassy dynamics. This motivated the search for direct entropy–dynamics relations. The most influential step in this direction was the Adam–Gibbs theory \cite{10.1063/1.1696442}, which proposed that structural relaxation requires cooperative rearrangements involving a minimum number of particles, with this number governed by the configurational entropy \(S_c\). Their argument yields the well-known Adam–Gibbs relation for the temperature dependence of the relaxation time \(\tau(T)\), given by \(\tau(T)= \tau_0 e^{C/TS_c}\), where \(\tau_0\) and \(C\) are material-specific constants. This relation has been widely applied to supercooled liquids near the glass transition temperature, where the entropy crisis drives both the dramatic slowing down of relaxation and the concomitant decrease in diffusion. 

A different relation between diffusion and entropy was proposed by Rosenfeld \cite{YaakovRosenfeld_1999} who observed numerically that a scaled (by microscopic parameters) diffusion coefficient \(D^*\) in liquids could be fitted rather well by a relation of the form \(D^* = a e^{bS_{ex}}\), where \(a\) and \(b\) are fitting parameters, while the excess entropy \(S_{ex}= S- S_{id}\) is the difference between the entropy of liquid under given condition and that of the ideal gas state of the same liquid. This scaling relation, known as Rosenfeld scaling, has been extensively used and tested \cite{10.1063/1.5055064,10.1063/1.2140282,10.1063/1.4967939,10.1063/1.4706520,BANERJEE2017}, and found to be semi-quantitatively reliable. Together with Dzugutov’s universal scaling law \cite{Dzugutov1996}, these scaling relations further reinforce the idea that entropy can serve as a unifying descriptor of transport and relaxation, connecting equilibrium thermodynamics with dynamical behavior across a wide range of liquids.

While many studies have examined both the diffusion-entropy scaling \cite{acharya2024relationentropydiffusionrelaxation,10.1063/5.0022818,10.1063/5.0116299,10.1063/5.0254388} and the influence of dimensionality on diffusion \cite{10.1063/1.4948936}, the specific interest in the dimensional dependence of diffusion-entropy scaling is a recent development \cite{PhysRevLett.131.147101,liao2021new}. Notably, the theoretical works of Sorkin et al. \cite{PhysRevLett.131.147101} and Liao et al. \cite{liao2021new} both demonstrate an exponential dependence of the diffusion coefficient \(D\) on entropy \(S\), with the prefactors \(2/d\) and \(1/d\) respectively. Sorkin et al. \cite{PhysRevLett.131.147101} employed a stochastic approach, deriving a generalized \(D-S\) relation by expressing entropy in terms of the Shannon entropy of the steady-state configuration distribution, which itself depends on the diffusion coefficient. In contrast, Liao et al. \cite{liao2021new} derived an effective Hamiltonian for a Brownian particle in a heat bath under the weak-coupling limit, replacing the full microscopic Hamiltonian. This allowed for the analytical computation of the partition function, from which the \(D-S\) scaling relation was obtained. 

Importantly, Sorkin et al. \cite{PhysRevLett.131.147101} also established a two-state diffusion–entropy relation, enabling the entropy difference between two distinct states to be expressed in terms of their respective diffusion coefficients. Such relations are particularly valuable, as they provide a practical route to infer entropy changes from diffusion measurements. However, the validity of these theoretical relations remains to be rigorously tested through experiments and computer simulations.

In this study, we examine the dimensional dependence of the relationship between diffusion coefficients and entropies of two distinct thermodynamic states of a system. In Section \ref{sec:2}, we briefly review recent theoretical works \cite{PhysRevLett.131.147101,liao2021new} that have explored how dimensionality influences diffusion–entropy scaling. We also present a theoretical derivation of the two-state diffusion–entropy relation for the specific case of a particle undergoing Brownian motion in \(d\) dimensions. Our calculation supports a dimensional prefactor of \(2/d\) consistent with earlier findings. In Section \ref{sec:3}, we turn to molecular dynamics (MD) simulations to test the validity of this relation. Specifically, we assess the validity of this relationship for two distinct systems: Lennard-Jones (LJ) fluid (an atomic fluid) and water (a molecular fluid). Two-phase thermodynamic (2PT) \cite{10.1063/1.1624057,doi:10.1021/jp103120q,PannirSivajothi2019} method is used to calculate the entropy of the given systems. Details of the simulation methodology are presented in section \ref{sec:3a}.

Our results, presented in section \ref{sec:3b}, show that the ratio of diffusion coefficients of two distinct thermodynamic states follows an exponential dependence on their entropy difference, regardless of the system considered. In 2\(d\) and 3\(d\) LJ systems, we see a clear dimensional dependence where the prefactor \(\alpha/d\) before the entropy difference \(\Delta S\) (eqn \ref{eq:3}) has an intermediate value between \(1/d\) and \(2/d\). As water is a molecular fluid, the validity of the above relation has been studied while considering both total and translational entropy difference between the states. However, for water, the values of \(\alpha/d\) (evaluated for both total and translational entropy) show little variation with dimension \(d\) and are closer to the Rosenfeld scaling coefficient \cite{10.1063/1.4967939} of water. Furthermore, we demonstrate that the above exponential dependence is also obeyed even when only the rotational entropy of water is considered, although the values of \(\alpha/d\) are substantially higher than their translational counterparts. Finally, in section \ref{sec:4}, we conclude by discussing the possible reasons for the discrepancy in the diffusion-entropy relationship of LJ liquid and water.
\section{Theoretical Analysis}
\subsection{\label{sec:2}Prior work on dimensional dependence of \(D-S\) relation}
While diffusion–entropy scaling has been widely studied across a range of systems, the role of spatial dimensionality in shaping this relation has only recently come under theoretical scrutiny \cite{PhysRevLett.131.147101,liao2021new}. Earlier studies, such as Rosenfeld’s \cite{YaakovRosenfeld_1999}, established empirical exponential relationships between a scaled diffusion coefficient and excess entropy but lacked explicit dimensional dependence. In contrast, recent theoretical frameworks based on stochastic dynamics and effective Hamiltonian representations have provided analytical justification for a dimension-dependent form of the diffusion–entropy scaling.  
 
These theoretical works employed a stochastic framework, typically using solutions of the diffusion or Fokker-Planck equations for the probability density \(P(\vec{x},t)\) of a particle's position \(\vec{x}\) at time \(t\), to establish a relationship between diffusion and entropy from the perspective of a diffusing particle. Diffusion coefficient \(D\) describes the time evolution of the probability distribution function, and the entropy is obtained from the logarithm of the distribution function, so the quantities are intimately connected. This stochastic approach should be reliable unless the underlying relaxation demands a more complex probability distribution function.

 Specifically, Sorkin \textit{et al.} \cite{PhysRevLett.131.147101} demonstrated, using this stochastic formulation, that the relationship between diffusion and entropy exhibits a strong dependence on the spatial dimensionality \(d\), arising from the form of the propagator that governs time evolution from an initial state. Assuming Markovian dynamics and weak mixing of particle trajectories, the configuration probability distribution is expressed as a product of single-particle propagators and the initial configuration distribution. If the single-particle propagator converges to a Gaussian for times greater than the characteristic relaxation time \(\tau\), their theory predicts that the ratio of diffusion coefficients corresponding to two thermodynamic states (denoted 1 and 2) satisfies an exponential relation with their entropy difference:

\begin{equation}
D_1 \tau_1 \geq D_2 \tau_2 \exp\left( \frac{2}{d} \Delta S \right),
\label{eq:1}
\end{equation}

where \(\Delta S = (S_1-S_2)/k_B\) is the difference in entropy per particle between the two states and \(k_B\) is the Boltzmann constant. The dimensionality-dependent factor \(2/d\) originates from the prefactor of the single particle propagator. This inequality becomes an equality in the special case where particle trajectories do not mix and the system's dynamics are fully captured by independent single-particle diffusion.

 Another recent theoretical treatment \cite{liao2021new} by Liao \textit{et al.} also demonstrated the dimensional dependence of diffusion entropy scaling. They employed an earlier work of mathematician Bateman \cite{PhysRev.38.815}, who explored the possibility of replacing, for certain purposes, a dissipative non-conservative system by a conservative Hamiltonian system. We are aware of the reverse process where a solute coupled to a bunch of harmonic oscillators can be replaced by a generalized Langevin equation as demonstrated by Zwanzig \cite{Zwanzig1973}. Bateman \cite{PhysRev.38.815} earlier had shown that it would be possible to replace a Langevin equation by a Hamiltonian system such that both the systems followed the same phase space trajectory once the initial conditions are specified. Liao and Gong \cite{liao2021new} leveraged this dynamic equivalence to derive an effective Hamiltonian corresponding to the Langevin dynamics when viewed in the reference frame of a Brownian particle at equilibrium. Concretely, starting from the Langevin equation \(m\ddot{x} + \gamma \dot{x} = \zeta(t) ,\) (where $m$ is the particle mass, $\gamma$ the friction coefficient, and $\zeta(t)$ the stochastic force with $\langle \zeta(t)\rangle=0$),  they shift to the particle’s frame and introduce the relative coordinate \(y=x-x_o\). This yields a second-order deterministic equation:
 \begin{equation}\label{eq:li1}
m\ddot{y} + m\gamma \dot{y} = 0 .
\end{equation}
which can be equivalently generated by a harmonic potential,
\begin{equation}\label{eq:li2}
\phi(y) = \text{const} + m\gamma \nu_0\, y - \frac{1}{2} m \gamma^2\, y^2 ,
\end{equation}
where \(\nu_0\) is the initial velocity. The corresponding equilibrium effective Hamiltonian for an ensemble of $n$ particles is
\begin{equation}\label{eq:li2}
H_{\text{total}} = \sum_{i=1}^n \left[ \frac{p_i^2}{2m} + \frac{1}{2} m \gamma^2 (x_i - x_o)^2 \right] + \text{constant} .
\end{equation}

 Subsequent steps are straightforward and involve the calculation of entropy through the partition function via the free energy. This approach establishes a relation between diffusion and entropy, which can be expressed as
\begin{equation}
  D = \frac{\hbar}{e m} \exp\left[\frac{S}{k_B d}\right] .
  \label{eq:2}    
\end{equation}
Although put in Rosenfeld scaling form, the exponent with entropy has a dimensionality dependence of the form \(1/d\).

\subsection{ \(D-S\) Relation from Brownian Motion}
If one wants to compare the diffusion constants of two different thermodynamic states of the same system,
one needs to take cognition of the fact that the short-time dynamics in the two states can be quite different \cite{PhysRevLett.73.360}. The short-time dynamics depend explicitly on the nearest-neighbor arrangements, which are sensitive to the temperature and the density. One does not expect any kind of universality in the short-time dynamics. However, the time beyond which the diffusive motion of a tagged particle sets in could prove useful because one can adopt a diffusive dynamics and a Green's function description only beyond this time. Here, we can consider Shannon entropy as a measure of the disorder exemplified by the probability distribution. But we do need to remove the short-term contributions. It is easy in simulations, and probably also in theory. Several D-S scaling hypotheses/conjectures endorse this view.

An interesting question is whether such a description can be extended to two completely different systems, specifically, whether the diffusion coefficients of two distinct systems can be related through their entropy differences. Their short-time dynamics are clearly quite different, yet it remains to be explored whether entropy differences alone can still account for the observed differences in diffusion. This question is not addressed in the present work, but it would be worthwhile to explore in the future.

With this perspective in mind, it is instructive to revisit the problem from the simplest setting of single-particle Brownian motion. This allows us to derive the entropy–diffusion relation explicitly from the probability distribution and to assess how dimensionality enters the \(D-S\) relation.

Inspired by the work of Sorkin \textit{et al.} \cite{PhysRevLett.131.147101}, which relates steady-state configurational entropy to diffusion in many-particle systems, we consider an analogous relation for a single particle undergoing Brownian motion in d-dimensions. In the overdamped limit, the probability distribution \( P(\vec{x}, t)\) of the particle's position \(\vec{x}\) evolves according to the diffusion equation \cite{10.1093/oso/9780195140187.001.0001,bagchi2023nonequilibrium}: 
\begin{equation}\label{eq:a}
\frac{\partial P(\vec{x}, t)}{\partial t} = D \nabla^2 P(\vec{x}, t)
\end{equation}
The Green's function solution of the diffusion equation \ref{eq:a} in d-dimensions is,
\begin{equation}\label{eq:b}
P(\vec{x}, t) = \frac{1}{(4\pi D t)^{d/2}} \exp\left( -\frac{|\vec{x}|^2}{4Dt}\right)
\end{equation}
which represents a normalized Gaussian distribution. Now, we derive a relation between diffusion coefficient \(D\) and entropy \(S\) of the system by using the definition of Shannon entropy as follows:
\begin{equation}\label{eq:c}
S(t)=-k_B \int P(\vec{x},t) \mathrm{ln} P(\vec{x}, t) d^dx
\end{equation}
where \(k_B\) is the Boltzmann constant. Substituting the expression for \( P(\vec{x}, t)\) in equation \ref{eq:c}, and using the well known relation for mean square displacement \(<|\vec{x}|^2>=2dDt\), we obtain:
\begin{equation}\label{eq:d}
 S(t)/k_B=\frac{d}{2}\mathrm{ln}(4 \pi D t)+ \frac{d}{2}
\end{equation}
 For two particles with diffusion coefficients \(D_1\) and  \(D_2\), and entropies \(S_1(t)\) and \(S_2(t)\) at the same time t, we find:
 \begin{equation}\label{eq:e}
    \frac{D_1}{D_2}=\exp\left[\frac{2}{d}  \left(\frac{S_1-S_2}{k_B}\right)\right]
\end{equation}
Equation \ref{eq:e} demonstrates that the ratio of diffusion coefficients between two states depends exponentially on the difference in their entropies, scaled by the spatial dimension \(d\).

While the above derivation captures the entropy-diffusion relationship purely from positional dynamics, a more complete description of Brownian motion includes both position and velocity degrees of freedom. In this case, the dynamics are governed by the underdamped Langevin equation \cite{10.1093/oso/9780195140187.001.0001,bagchi2023nonequilibrium}, and the evolution of the joint probability distribution in phase space is described by the Kramers equation (in the absence of external potential). At long times, the joint probability distribution of phase space  \( P(\vec{x},\vec{v}, t)\) is given by:
\begin{multline}
\label{eq:f}
P(\vec{x},\vec{v}, t) = 
\frac{1}{(4\pi D t)^{d/2}} \exp\left( -\frac{|\vec{x}|^2}{4Dt} \right) \times \\ 
\left( \frac{m}{2\pi k_B T} \right)^{d/2} \exp\left( -\frac{m|\vec{v}|^2}{2k_B T} \right)
\end{multline}

Here, the velocity distribution has relaxed to the time-independent Maxwell-Boltzmann form \(P_{MB}(\vec{v})\), and position distribution \(P(\vec{x}, t)\) is diffusive, i.e, \(P(\vec{x},\vec{v}, t)=P(\vec{x}, t) \times P_{MB}(\vec{v})\). 
The total Shannon entropy \(S(t)\) is given by:
\begin{equation}\label{eq:g}
    S(t)=-k_B \int P(\vec{x},\vec{v,}t) \mathrm{ln} P(\vec{x},\vec{v}, t) d^dx d^dv
\end{equation}
Substituting equation \ref{eq:f} into the above, we see that the total entropy decomposes into independent contributions from position and velocity: \(S(t)=S_{pos}+S_{vel}\). The positional component of entropy \(S_{pos}\) is given by equation \ref{eq:d}. Analogous expression for the velocity component of entropy \(S_{vel}\) is obtained by using equipartition theorem  \(<|\vec{v}|^2>=dk_BT/m\). The final expression for total entropy \(S(t)\) is given below:
\begin{equation}\label{eq:h}
S(t)/k_B=\frac{d}{2}\mathrm{ln}(4 \pi D t)+ \frac{d}{2}\mathrm{ln}\left( \frac{2 \pi k_BT}{m}\right)+d
\end{equation}
Reframing the above equation in the two-state form we get,
\begin{equation}\label{eq:i}
    \frac{D_1T_1}{D_2T_2}=\exp\left[\frac{2}{d}  \left(\frac{S_1-S_2}{k_B}\right)\right]
\end{equation}
We observe that the prefactor \(2/d\) in the entropy–diffusion relation remains unchanged even after including the velocity distribution. This is not entirely surprising, as both the position and velocity probability distributions retain Gaussian form, and their entropies contribute additively. However, an important distinction arises: the inclusion of velocity introduces an explicit dependence on temperature \(T\) in the \(D-S\) relation. 

Our derivation of the \(D-S\) relation applies specifically to the special case of single-particle Brownian motion and cannot be directly generalized to many-body interacting systems. In contrast, the theoretical frameworks proposed by Sorkin \cite{PhysRevLett.131.147101} \textit{et al.} and Liao \cite{liao2021new} \textit{et al.} are more general in scope but still rely on certain approximations. Similarly, Rosenfeld scaling is fundamentally phenomenological. It can, of course, be derived (as shown in reference \cite{bagchi2013water}) via the assumption of a random walk where the rate of transition between regions in phase space is proportional to the density of microscopic states, but the derivation also involves serious approximations. Given that all current theoretical approaches to \(D-S\) scaling involve approximations, it is essential to assess their validity across different systems through experimental measurements or computational studies. 
It is convenient to consider the above relations in the form of the ratio of diffusion coefficients between two distinct thermodynamic states, designated here by 1 and 2,
\begin{equation}
  D_{1}/D_{2} = \exp\left[\frac{\alpha}{d}\left(\frac{S_1-S_2}{k_{B}}\right)\right]
  \label{eq:3}
\end{equation}
where \(S_1\) and \(S_2\) are the entropies per particle of thermodynamic states 1 and 2, respectively.
The value of prefactor \(\alpha/d\), which includes the dimensionality \(d\) of the system, can easily be studied by computer simulations as shown in section \ref{sec:3}. Thus, the dimensionality dependence can be evaluated, allowing us to validate theoretical predictions.

\begin{figure}[htp]
\centering
\begin{minipage}{0.5\textwidth}
  \includegraphics[width=0.9\textwidth, height=5cm]{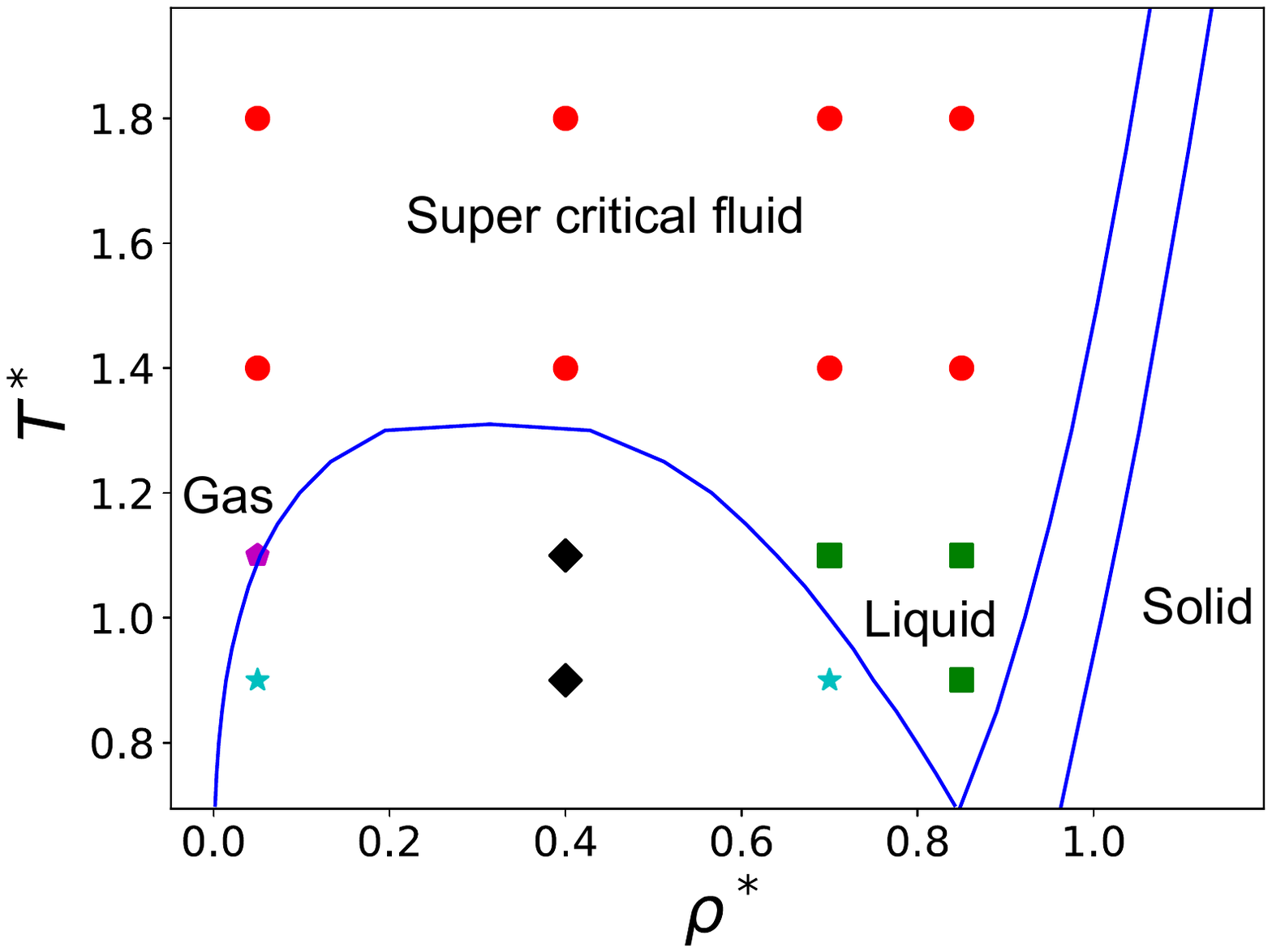}

\end{minipage}%
\caption{The figure presents the phase diagram of the \(3d\) LJ system in density \(\rho^*\) -- temperature \(T^*\) phase space. Entropy and diffusion coefficients were evaluated at 16 distinct thermodynamic state points, which are indicated on the diagram. These state points span various phases, including gas (magenta pentagons), liquid (green squares), supercritical fluid (red circles), unstable states (black diamonds), and metastable states (cyan stars).}
\label{fig:0}
\end{figure}
\section{Numerical analysis} \label{sec:3}
\subsection{Methodology and computational details} \label{sec:3a}
\subsubsection{Molecular dynamics (MD) simulations}
We perform molecular dynamics (MD) simulations of the Lennard-Jones (LJ) system in both \(2d\) and \(3d\). We also conduct MD simulations of liquid water under various conditions: bulk (\(3d\)), confined inside a graphene slit pore (\(2d\)), and confined within carbon nanotube (CNT) (\(1d\)).

 \textit{LJ liquid:} The interaction between the LJ particles is given by the Lennard-Jones potential given below:
 \begin{equation}
    U_{LJ}(r)=4\epsilon\bigg[\bigg(\frac{\sigma}{r}\bigg)^{12}-\bigg(\frac{\sigma}{r}\bigg)^6\bigg]
\label{eqn:LJ}    
\end{equation}
Here, \(\epsilon\) is the strength of the interaction, \(\sigma\) is the diameter of the LJ particle, and \(r\) is the distance between the interacting particles. We have used the parameters of argon (\(\sigma\)= 3.405 Å, \(\epsilon\) = 0.238 kcal  mol\textsuperscript{-1} and
mass m = 39.948 g  mol\textsuperscript{-1}) in the MD simulations. However, the results for LJ systems are presented in reduced units, where we take \(\sigma\), \(\epsilon\), and the mass \(m\) of the particles as units of length, energy, and mass. Hence, the density and temperature in reduced units are given by \(\rho^*=\rho \sigma^3\) (3d) or \(\rho \sigma^2\) (2d) and \(T^*=k_BT/\epsilon\). Simulations were conducted in the NVT ensemble, and after equilibrating, trajectories of 20 ps were saved at a frequency of 2 fs for entropy calculation at each state point. The Nosé-Hoover \cite{evans1985nose} thermostat was used to maintain the temperature of the system with a time constant of 0.05 ps.

 The \(3d\) simulations were performed at 16 thermodynamic state points \cite{10.1063/1.1624057} in the phase diagram with density \(\rho^*\) varying from \(0.05\) to \(0.85\) and temperature \(T^*\) ranging from \(0.9\) to \(1.8\).  In \(2d\), simulations were performed at 12 thermodynamic state points \cite{PannirSivajothi2019} with the density and temperature in the range of \(0.01 \leq \rho^* \leq 0.77 \) and \(0.45 \leq T^* \leq 1\) respectively. The state points of the \(3d\) LJ system considered here, along with the corresponding phase diagram, are illustrated in Fig. \ref{fig:0}. These states span different phases in the phase diagram, ranging from gas to liquid, and include unstable and metastable states (excluding solid states). The readers are referred to references \cite{10.1063/1.1624057} and \cite{PannirSivajothi2019} for further details on LJ systems.

\textit{Water:} Atomistic MD simulations are performed for water in bulk and confinement using the flexible water model TIP4P-2005f \cite{10.1063/1.2121687}. The simulations were conducted for bulk water in the liquid phase under periodic boundary conditions and will be referred to as \(3d\) water. \(2d\) water simulations were performed by confining liquid water inside a graphene slit pore, and \(1d\) confinement was achieved by confining water inside a carbon nanotube (CNT) of chirality (11,11). Starting at an initial temperature of 300 K, three-dimensional water was sequentially cooled to 5 K while maintaining a constant pressure of 1 bar. NPT simulations were conducted for 40 ns at 28 thermodynamic states across the temperature range of 300 K to 5 K. This was followed by 10 ns of NVT simulations and an additional 100 ps of NVT simulations saved at 2 fs frequency for entropy calculations. Similar procedure was followed for 37 thermodynamic states of \(2d\) and \(1d\) water spanning a temperature range from 400 K to 10 K. The velocity-rescale \cite{10.1063/1.2121687} thermostat and Parrinello–Rahman barostat were used in NPT simulation with 1.3 ps and 2.3 ps coupling constants, respectively. All simulations were performed using open-source LAMMPS \cite{LAMMPS} software. Readers are referred to reference \cite{10.1063/5.0047656} for further details on water simulations.
\subsubsection{Entropy calculation}
We have used the Two-phase thermodynamic (2PT) method to calculate the entropy and diffusion coefficient of LJ particles and water. The 2PT method \cite{10.1063/1.1624057,doi:10.1021/jp103120q,PannirSivajothi2019} provides an accurate and efficient means to calculate the thermodynamic properties of a system, like entropy and free energy, using short runs of MD simulations. The 2PT method has been successfully used to calculate the entropy of different molecular systems \cite{doi:10.1021/jp103120q,C2CP42011B,Huang2011} using trajectories as short as 20 ps. In our present work, we use 2PT method to obtain entropies of the thermodynamic states of \(2d\) and \(3d\) LJ system and \(1d\), \(2d\), and \(3d\) liquid water.

In the 2PT method, the density of states (DoS) of a system, which includes the normal modes associated with translation, rotation, and intermolecular vibration, is obtained by the Fourier transform of the velocity autocorrelation function. Trajectories of length as short as 20 ps dumped at a frequency of 2 fs are used to calculate the velocity auto-correlation function. A fluidicity factor is calculated by taking the ratio of the diffusivity of the system to that of the hard sphere gas at the same temperature and density. Subsequently, the fluidicity factor is used to decompose the density of states into non-diffusive solid-like and diffusive gas-like components. Thermodynamic properties, such as entropy \(S\), can be determined by integrating the density of states (DoS) with appropriate weighting functions, obtained by applying quantum statistics of harmonic oscillator for the solid components, hard sphere statistics for the translational gas component, and rigid rotor statistics for the rotational gas component. The additional details of the 2PT method are elaborately elucidated in references \cite{doi:10.1021/jp103120q,10.1063/1.1624057}. The diffusion coefficient \(D\) is calculated from the zeroth frequency of the density of states (DoS) by using the relation given below.
\begin{equation}
   DoS(0)=\frac{4dmND}{k_BT} 
\end{equation}
where \(m\) and \(N\) are the mass and number of particles, respectively, and T is the temperature of the system.
\\ \\
\begin{figure}[htp]
\centering
\begin{minipage}{0.5\textwidth}
  \includegraphics[width=0.7\textwidth, height=4cm]{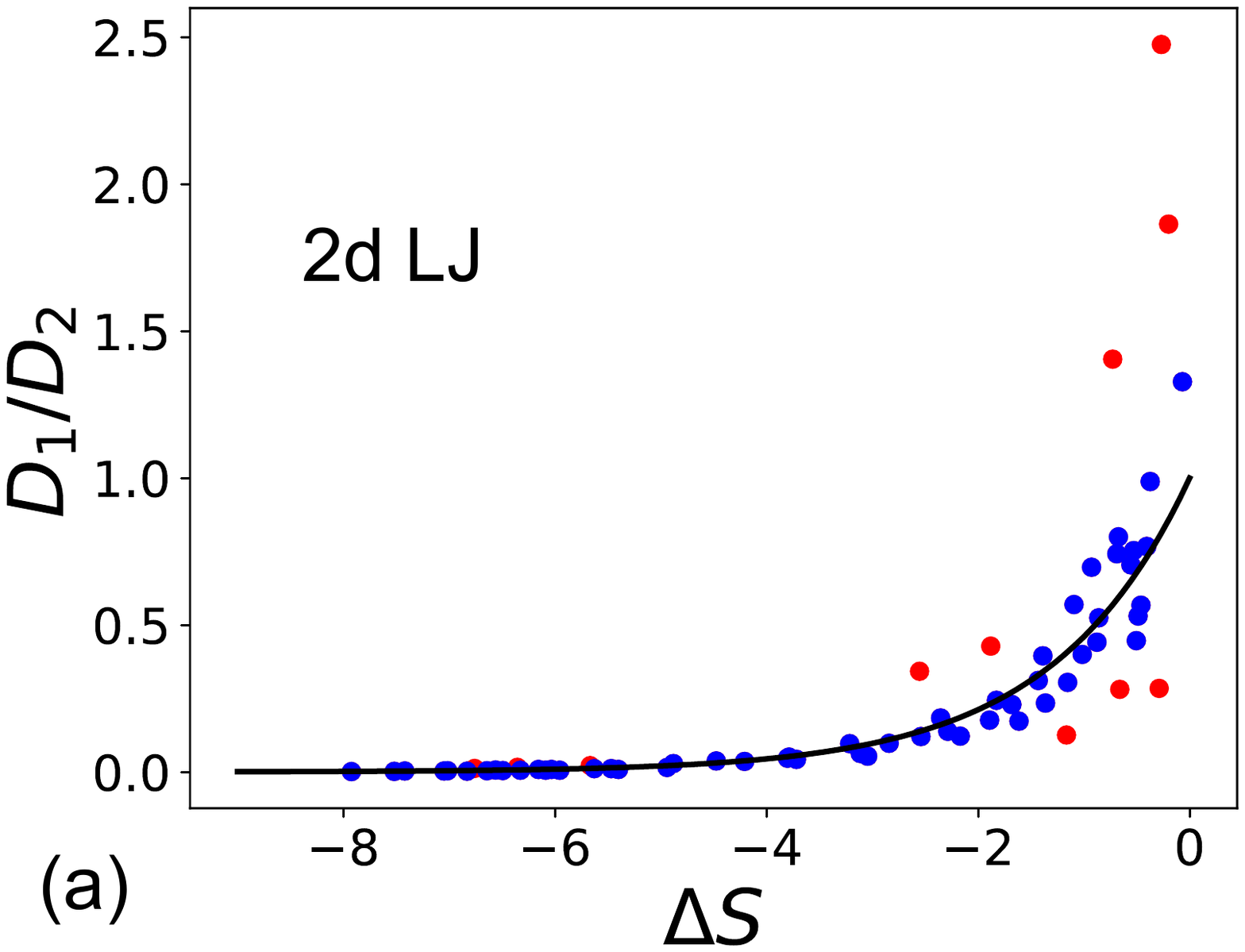}
  \includegraphics[width=0.7\textwidth,height=4cm]{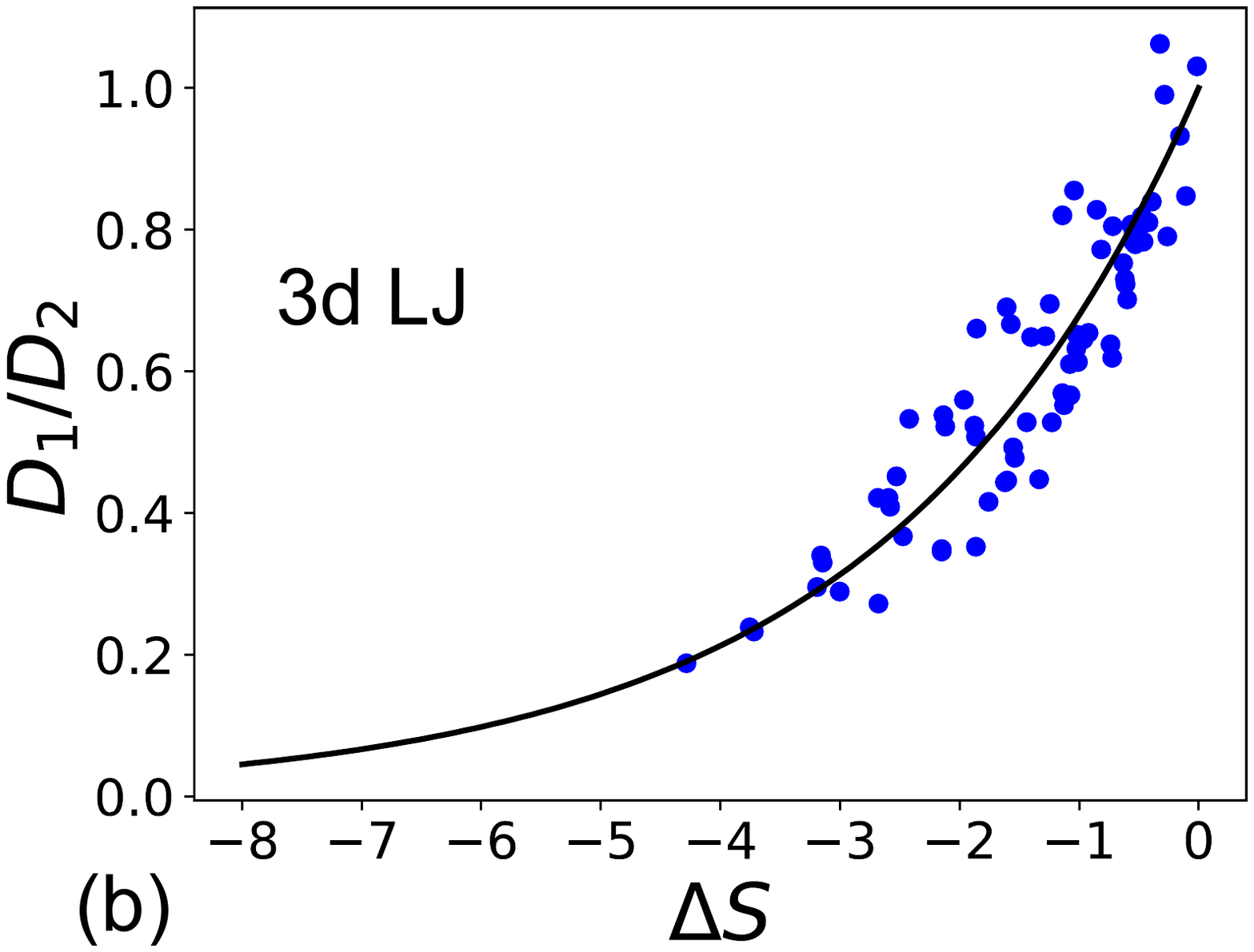}
\end{minipage}%
\caption{The figure illustrates the two-state diffusion entropy relation for \(2d\) and \(3d\) LJ systems, respectively. We plot the ratio of diffusion coefficients \(D_1/D_2\) between states 1 and 2 as a function of their entropy difference \(\Delta S\). The data points of the plot are given in blue dots. The red dots in \(2d\) LJ system (fig (a)) belong to the thermodynamic state pairs, where one of the thermodynamic states is unstable. The black line represents the data fitting curve of the form \(D_1/D_2=\exp[(\alpha/d) \Delta S]\). From the curve fitting, the value of \(\alpha/d\) is 0.77 and 0.38 for \(2d\) and \(3d\) system respectively.}
\label{fig:1}

\end{figure}
\subsection{Results} \label{sec:3b}
\subsubsection{LJ Fluid}
Figures \ref{fig:1} (a) and (b) depict the plots of the ratio of diffusion coefficients \(D_1/D_2\) of two distinct thermodynamic states 1 and 2, as a function of the entropy difference between the two states \(\Delta S= (S_1-S_2)/k_B\) (represented by blue dots), for \(2d\) and \(3d\) LJ system respectively.  The plot of \(2d\) LJ (figure \ref{fig:1}(a)) system demonstrates that the ratio of diffusion coefficients \(D_1/D_2\) follows an exponential dependence on the entropy difference \(\Delta S\). However, as \(\Delta S\) approaches \(0\), a few data points show significant deviations from the exponential trend. On closer examination of the corresponding thermodynamic states of the data points, we find that the thermodynamic state pairs, wherein one of the states is unstable (state point at \(\rho^*=0.3\) and \(T^*=0.45\), represented by red dots), show significant deviations from the exponential trend especially near \(\Delta S=0\). This deviation is expected as the diffusion exhibits high fluctuations in the unstable state. So we exclude these data points from further analysis. To obtain the prefactor \(\alpha/d\) (equation \ref{eq:3}), that provides the best approximation for the plot, we fit the data (excluding red dots) with a curve of the form \(\exp[(\alpha/d) \Delta S]\) and obtain the values of \(\alpha\) and \(\alpha/d\) approximately equal to \(1.54\) and \(0.77\) respectively. The root mean square error (RMSE) of curve fit for \(2d\) LJ is \(0.095\).

Similarly, in Figure \ref{fig:ljall}, we also plot the ratio of diffusion coefficients \(D_1/D_2\) versus entropy difference \(\Delta S\) for the \(3d\) LJ system. The results for \(3d\) are much more complex than \(2d\), with the data points of the plot grouping into two branches, both with exponential dependence on \(\Delta S\). Branch 1, which deviates from the diffusion-entropy relation described in equation \ref{eq:3}, is obtained for those thermodynamic state pairs where one of the states is an LJ gas with very low density (\(\rho^*=0.05\)). However, branch 2 of the plot conforms to the functional form of the diffusion-entropy relationship given in equation \ref{eq:3}. This branch will be exclusively used in further analysis and is shown in Figure \ref{fig:1}(b). From curve fitting of the scatter plot in Figure \ref{fig:1}(b) with an exponential form similar to the \(2d\) system, we obtain the values of \(\alpha\) and \(\alpha/d\) approximately equal to \(1.14\) and \(0.38\) respectively, with the associated curve-fit RMSE equal to \(0.083\). Therefore, for both \(2d\) and \(3d\) LJ systems, the values of the prefactor \(\alpha/d\),  have intermediate values between theoretically predicted values of \(1/d\) and \(2/d\). This finding highlights a clear dimensional dependence in the diffusion entropy scaling for LJ systems.

\begin{figure}[htp]
\centering
\begin{minipage}{0.5\textwidth}
  \includegraphics[width=0.7\textwidth, height=4cm]{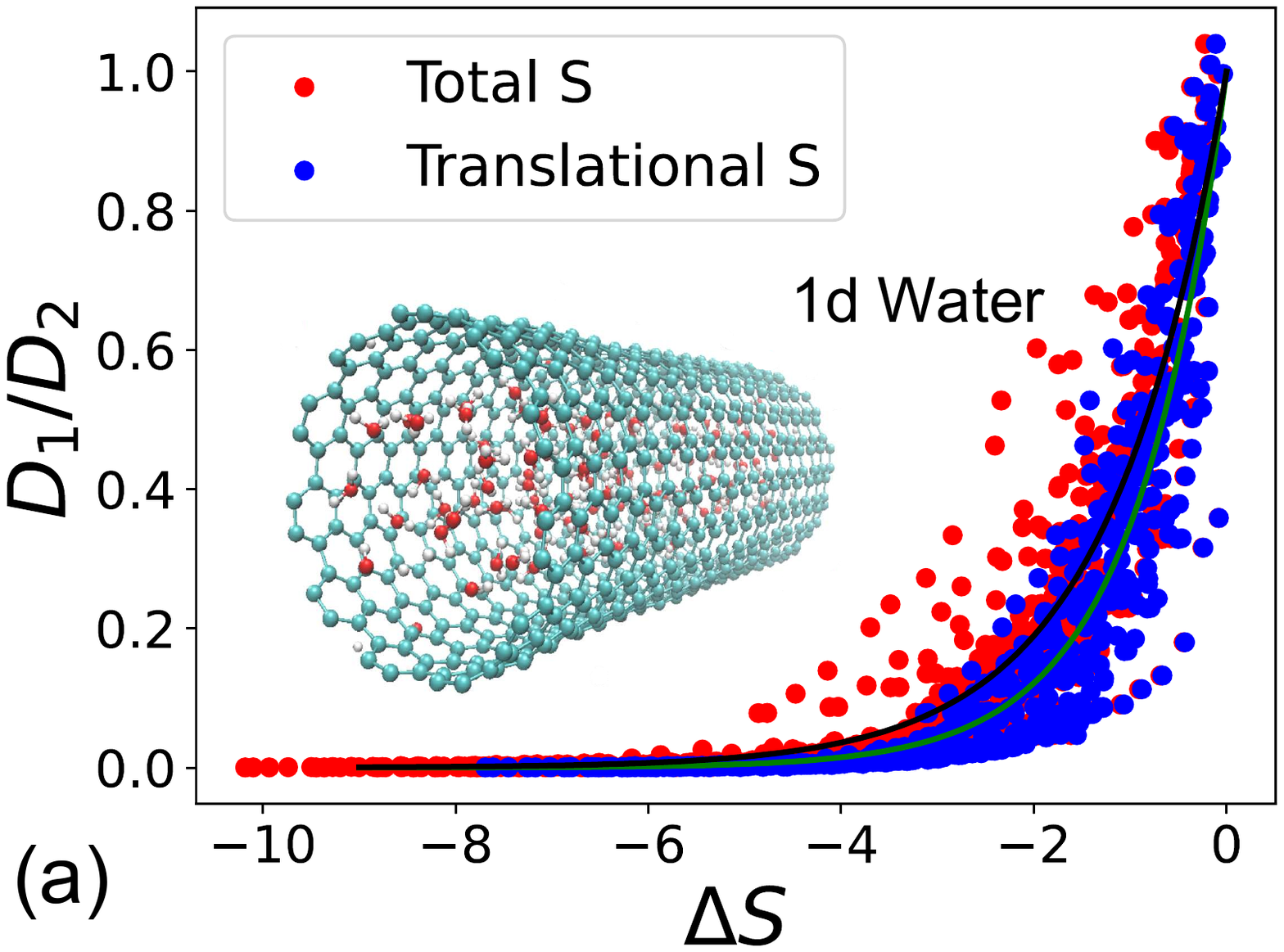}
  \includegraphics[width=0.7\textwidth, height=4cm]{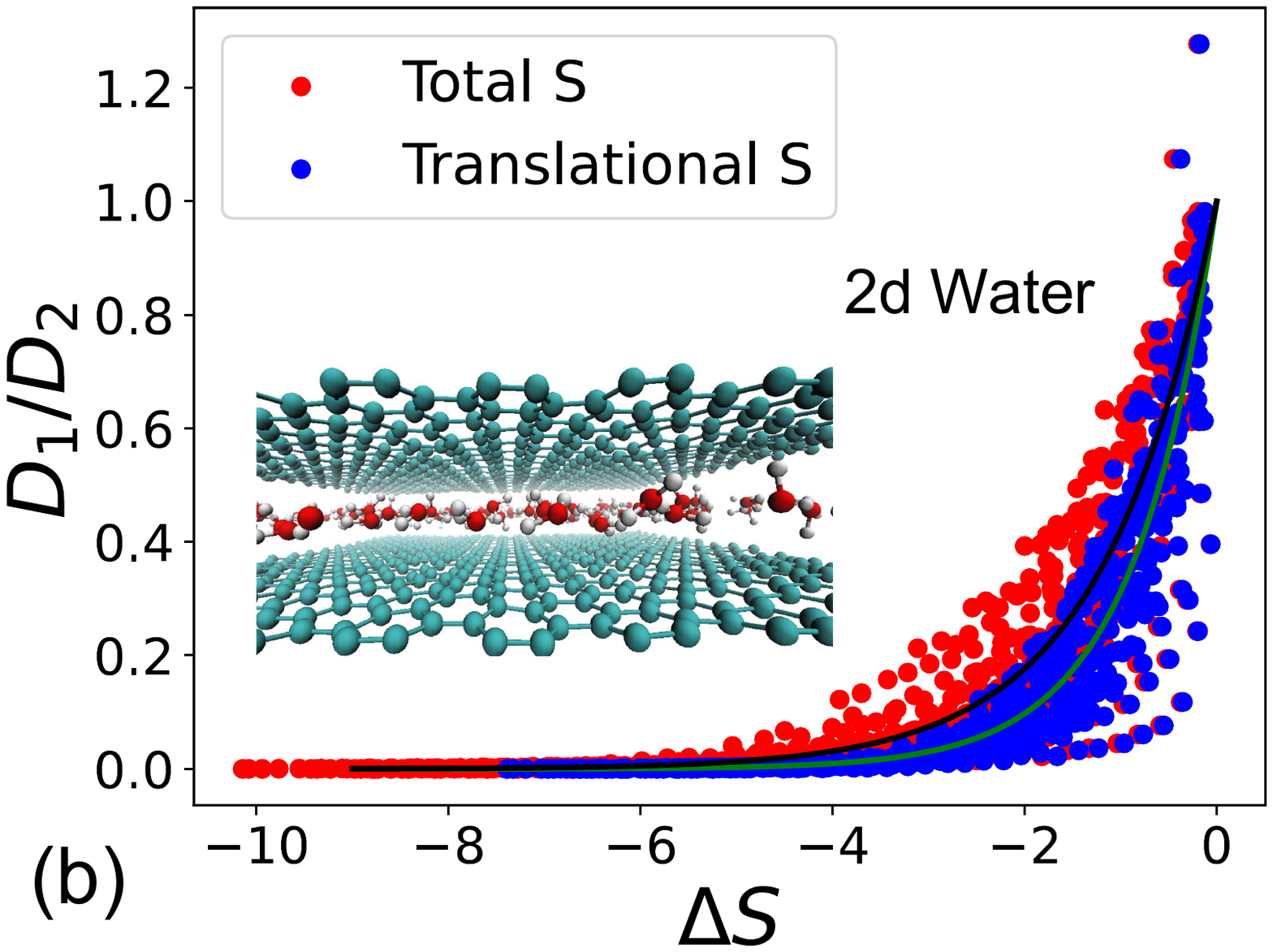}
  \includegraphics[width=0.7\textwidth, height=4cm]{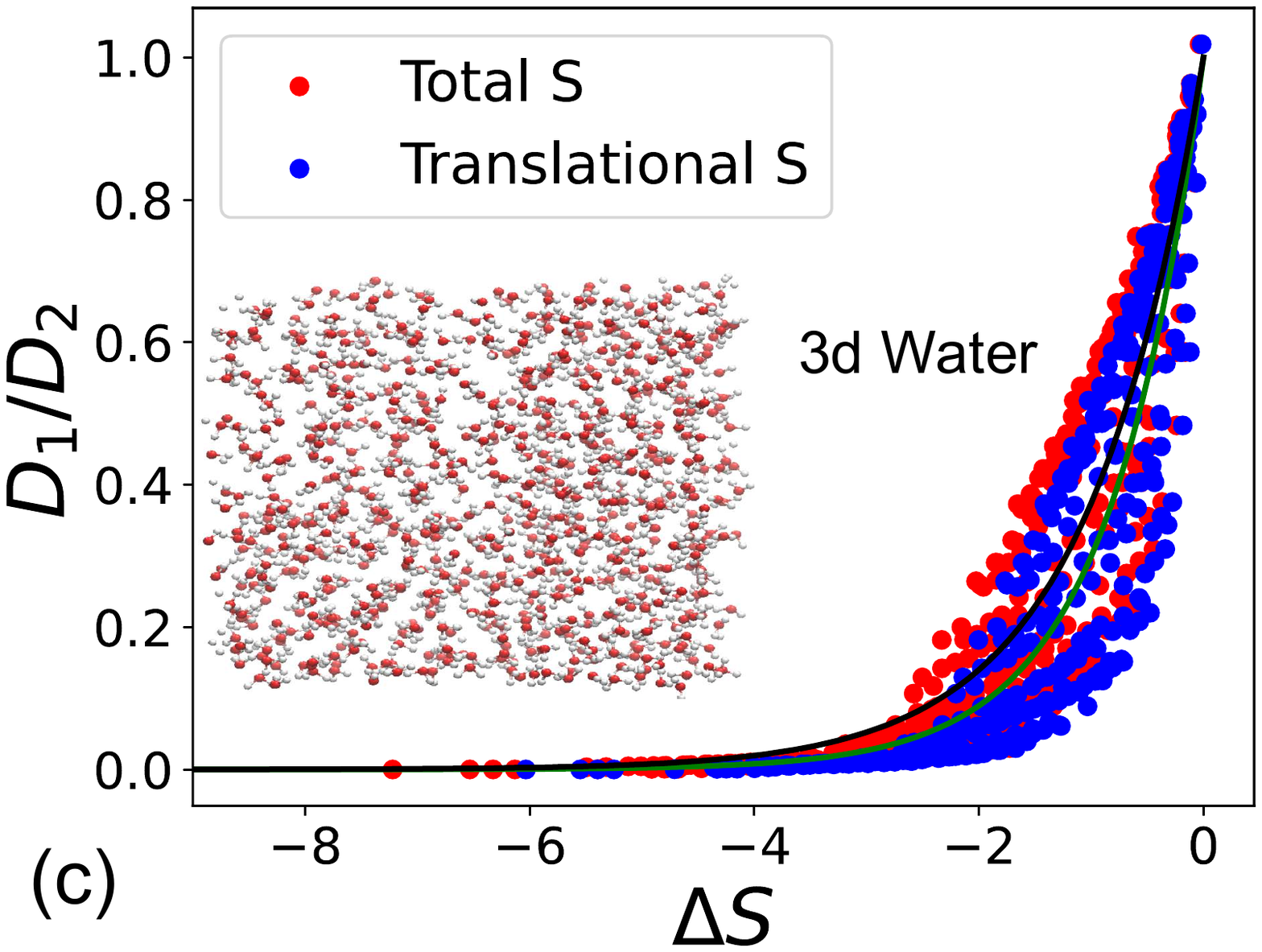}
  \label{fig:2}
\end{minipage}%

\caption{The figures (a),(b), and (c) illustrate the two-state diffusion entropy relation for \(1d\), \(2d\), and \(3d\) liquid water, respectively. The inset of the plots shows the instantaneous snapshots of water in their corresponding dimensions. The oxygen and hydrogen atoms in water molecules are represented by red and white colors, respectively. The carbon atoms in graphene (\(2d\)) and carbon nanotube (\(1d\)) are in blue color. The red scatter plot illustrates the ratio of diffusion coefficients \(D_1/D_2\) between states 1 and 2 as a function of their total entropy difference \(\Delta S\). The blue scatter plot shows \(D_1/D_2\) between states 1 and 2 as a function of their translational entropy difference. The black and green lines represent the data fitting curve of the form \(D_1/D_2=\exp[(\alpha/d) \Delta S]\) for total and translational entropy difference, respectively. For the case of total entropy difference (black curve), the value of \(\alpha/d\) is 0.83, 0.86, and 0.98 for 1\(d\), 2\(d\) and 3\(d\) systems, respectively. For the case of translational entropy difference (green curve), the value of \(\alpha/d\) is 1.05, 1.16, and 1.21 for \(1d\), \(2d\), and \(3d\) systems, respectively. }
 \label{fig:2}
\end{figure}
\subsubsection{Water}
Unlike the Lennard-Jones (LJ) system, water is a molecular fluid whose entropy arises from translational, rotational, and vibrational motions of its molecules. Therefore, we study the diffusion entropy relation using the total, translational, and rotational entropy of water. Furthermore, we highlight another significant distinction between the LJ system and water: in confined (\(2d\) and \(1d\)) scenarios, water is not perfectly constrained to a \(2d\) plane or a \(1d\) line. Instead, its motion is restricted by narrow slits or nanotubes (insets in Figure \ref{fig:2}), resulting in a small but non-zero entropy contribution from the direction perpendicular to the confinement.  Figure \ref{fig:2} (a), (b) and (c) illustrate the plots of ratio of diffusion coefficients \(D_1/D_2\) between two distinct thermodynamic states 1 and 2, as a function of the entropy difference \(\Delta S\) between the two states for \(1d\), \(2d\) and \(3d\) water respectively. The red data points correspond to the above relation based on the total entropy difference, while the blue points represent those based solely on the translational entropy difference. It is evident that the diffusion ratio \(D_1/D_2\) exhibits an exponential dependence on the entropy difference \(\Delta S\), regardless of the dimensionality of the water. Unlike the LJ system in 3D, no significant branching of data points is observed for water in any dimension. Instead, a notable dispersion in the distribution of data points is observed as \(\Delta S\) approaches zero. Fitting the data with a curve of the form \(D_1/D_2=\exp[(\alpha/d) \Delta S]\) (black curve for total entropy and green curve for translational entropy), we obtain the value of the prefactor \(\alpha\). The values of \(\alpha\) are approximately equal \(0.83\), \(1.72\) and \(2.94\) for \(1d\), \(2d\), and \(3d\) water respectively, in the case of \(\Delta S\) being the total entropy difference. When \(\alpha\) is divided by dimension \(d\), contrary to the LJ system, the values of \(\alpha/d\) show little variation with dimension. The computed values are approximately equal to 0.83, 0.86, and 0.98 for \(1d\), \(2d\), and \(3d\) water, respectively, all of which are close to Rosenfeld exponent of water \cite{10.1063/1.4967939} \(b \approx 1\). The associated RMSE of the curve fit for \(1d, 2d\) and \(3d\) water is 0.098, 0.1, and 0.11, respectively. 

When \(\Delta S\) accounts only for translational entropy difference, we obtain the value of \(\alpha\) approximately equal to 1.05, 2.32, and 3.63 for \(1d\), \(2d\), and \(3d\) water, respectively. The corresponding values of \(\alpha/d\) are \(1.05\), \(1.16\), and \(1.21\), showing minimal variation with \(d\), similar to the trend observed with total entropy.  The RMSE of the curve fit in the above case for \(1d\), \(2d\), and \(3d\) water is 0.087, 0.089, and 0.11, respectively. The plots for translational entropy are shifted right with respect to the total entropy plot, as the value of translational entropy is always less than total entropy.  As a result, the value of \(\alpha\) is higher for translational entropy than total entropy at all dimensions. Additionally, the deviation of data points from the fitting curve is lower in the case of translational entropy for \(1d\) and \(2d\) water. This suggests that as we exclude the contributions of rotational and vibrational motion from entropy, the diffusion-entropy relation shows enhanced adherence to the exponential form.

To further probe the role of rotational motion, Figure \ref{fig:rot} (Appendix) presents the exponential relation using only the rotational entropy difference. While the relation still holds, the exponential prefactor \(\alpha/d\) is substantially higher and approximately equal to \(5.39\), \(3.95\) and \(5.99\) for \(1d\), \(2d\) and \(3d\) water respectively. This unnaturally high value of the prefactor shows that the inherent dependence of rotational entropy on diffusion is weak. 

Overall, irrespective of whether the entropy in D-S relation is total entropy or translational entropy, the values of the pre-factor \(\alpha/d\) do not show a significant change with the dimension of water, revealing a weak dimensionality dependence of the two-state diffusion-entropy relation in water. The summary of the results discussed here is given in Table \ref{tab:my_label}.
\\ \\
\begin{table}[]
    \centering
    \begin{tabular}{|c|c|c|c|c|c|}
   \hline Dimension&LJ&Water& Water  &\(\alpha/d\)&\(\alpha/d\) \\
                   & & (total &(translational &\(=2/d\) &\(=1/d\) \\ 
                   & & entropy) & entropy) & &  \\ \hline
                   
 1d & -- & 0.83& 1.05& 2 & 1\\  \hline
 2d & 0.77 & 0.86 &1.16& 1 & 0.5\\ \hline
  3d & 0.38 &0.98 &1.21 & 0.66 &0.33\\ \hline

 \hline    
    \end{tabular}
    \caption{This table summarizes the value of prefactor \(\alpha/d\) in equation \ref{eq:3} obtained for LJ system and liquid water at different dimensions. The values of \(\alpha/d\) obtained by the theoretical work of Sorkin et al.\cite{PhysRevLett.131.147101}(equation \ref{eq:1}) and Liao et al.\cite{liao2021new} (equation \ref{eq:2}) are also included in the table. We can clearly see that \(\alpha/d\) for LJ system is intermediate between \(1/d\) and \(2/d\), showing strong dimensional dependence. However, \(\alpha/d\) of liquid water, for both total and translational entropy cases, remains relatively unchanged with dimension, revealing a weak dimensional dependence.  }
    \label{tab:my_label}
\end{table}
\section{Concluding Remarks}\label{sec:4}

The diffusion of water molecules in the liquid state has been a subject of enormous interest since the pioneering studies of Laage and Hynes \cite{doi:10.1126/science.1122154}, who demonstrated that rotational diffusion in water proceeds not by small, continuous angular displacements but by large-amplitude, intermittent jumps of typically about 60 degrees. In the Laage--Hynes picture, these jumps occur through exchanges of water molecules between the first hydration shell and the second-nearest shell, rendering the process inherently collective. Even under nanoscale confinement, such as within carbon nanotubes \cite{doi:10.1021/jp904099f,10.1063/1.2565806}, reorientational relaxation still proceeds via angular jumps, in this case involving the interchange of the two hydrogen atoms of a water molecule forming hydrogen bonds with the same neighbor.

Building on this framework, Laage and co-workers \cite{doi:10.1021/acs.jpclett.2c00825} recently established that translational diffusion of water is also strongly influenced by these large-amplitude reorientational jumps. Superimposed on the continuous translational motion of the molecular frame, these jump events provide bursts of displacement, which dominate long-time diffusion. Complementary work by Offei-Danso \textit{et al.} \cite{Offei-Danso2023} showed that such angular jumps occur collectively in bursts, while earlier studies by Singh \textit{et al.} \cite{C0CP02081H} revealed that the survival time of the Laage--Hynes exchange mechanism grows with decreasing temperature, giving rise to a correlation length that increases upon cooling.

The existence of this jump diffusion mechanism renders water qualitatively different from Lennard--Jones (LJ) liquids, where diffusion proceeds in a more continuous manner. As Laage \textit{et al.} \cite{doi:10.1021/acs.jpclett.2c00825} emphasized, a faithful description of diffusion in water may require two distinct diffusion constants to separately account for the continuous background motion and the intermittent jump contributions. This distinction provides a natural explanation for our key finding: the weak dimensionality \(d\) dependence of the diffusion--entropy \((D-S)\) relation in water. In LJ liquids, where motion is continuous, dimensionality strongly affects the available phase space and thus the prefactor of the exponential \(D-S\) scaling. By contrast, in water the collective jump mechanism dominates diffusion, effectively bypassing the geometric constraints imposed by dimensionality. As a result, the entropy change associated with rearrangements is captured similarly across different dimensions, leading to the near universality of the \(D-S\) scaling observed in our study.

Thus, the lack of dimensionality dependence in water is not an anomaly but a direct manifestation of the Laage--Hynes jump mechanism. More broadly, this suggests that deviations from dimensionality-sensitive transport may serve as a diagnostic for identifying jump-dominated diffusion in other complex liquids, and highlights the practical utility of entropy--diffusion scaling as a probe of underlying molecular mechanisms.

\section*{Acknowledgment}
N.V. thanks MoE, India, for the fellowship. P.K.M. acknowledges funding through ANRF (erstwhile SERB), IRHPA (No. IPA/2020/000034).

\section*{Data Availability Statement}

The data that support the findings of this article are available from the authors upon reasonable request.

\appendix

\section{Appendix}
\setcounter{figure}{0}
\renewcommand{\thefigure}{A\arabic{figure}}

\subsection{Additional figures:}
 \begin{figure}[h!]
 \centering 
 \includegraphics[width=0.4\textwidth]{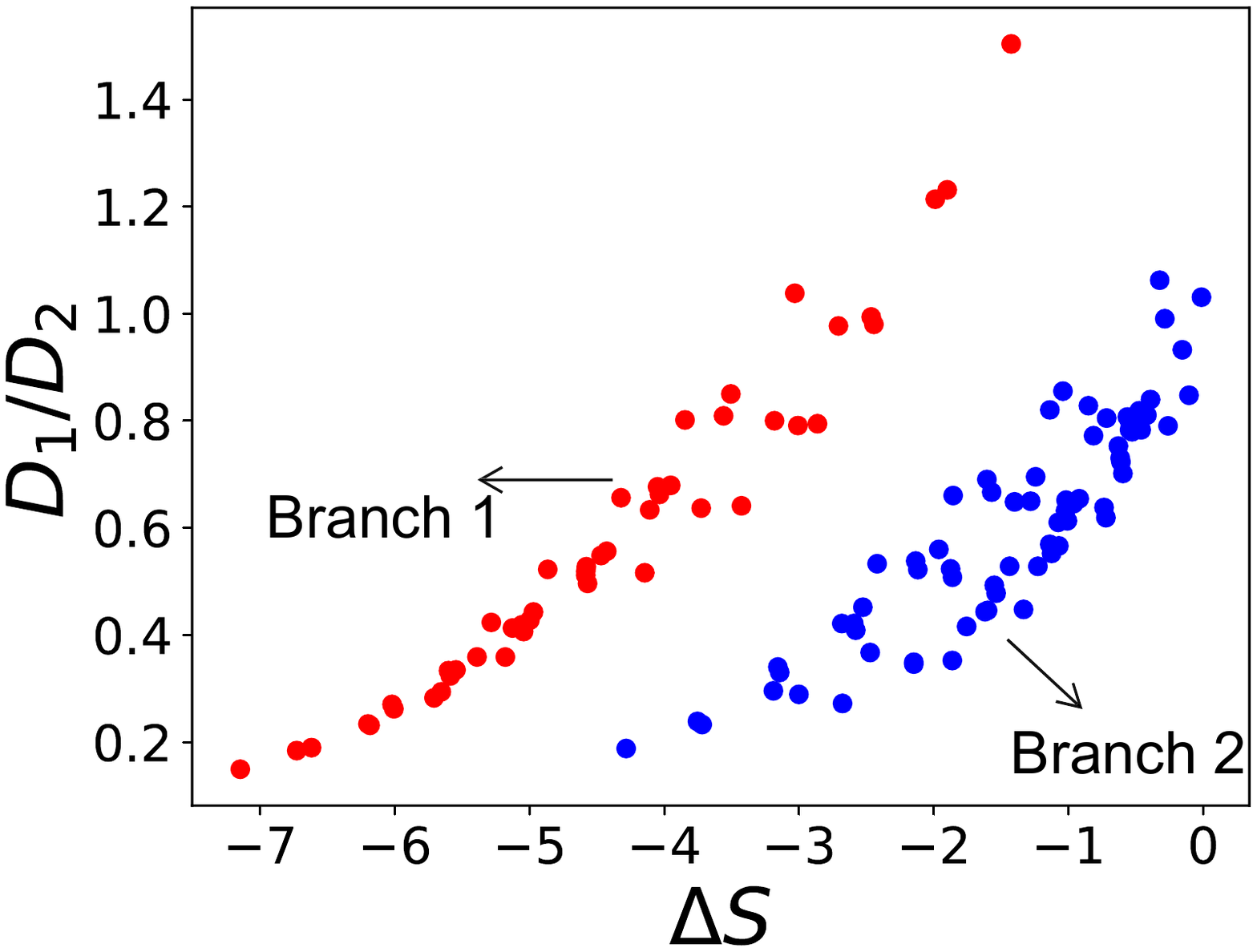}
 \caption{The figure illustrates the two-state diffusion entropy relation for \(3d\) LJ systems. We plot the ratio of diffusion coefficients between states 1 and 2 as a function of their entropy difference. The data points of the plot group into two branches(indicated by red and blue dots). Branch 1, which deviates from the diffusion-entropy relation described in equation \ref{eq:3}, is actually obtained for those thermodynamic state pairs where one of the states is an LJ gas with very low density (\(\rho^*=0.05)\).  }
 \label{fig:ljall}
 \end{figure}
\begin{figure}[h!]
\centering
\begin{minipage}{0.5\textwidth}
  \includegraphics[width=0.7\textwidth, height=4cm]{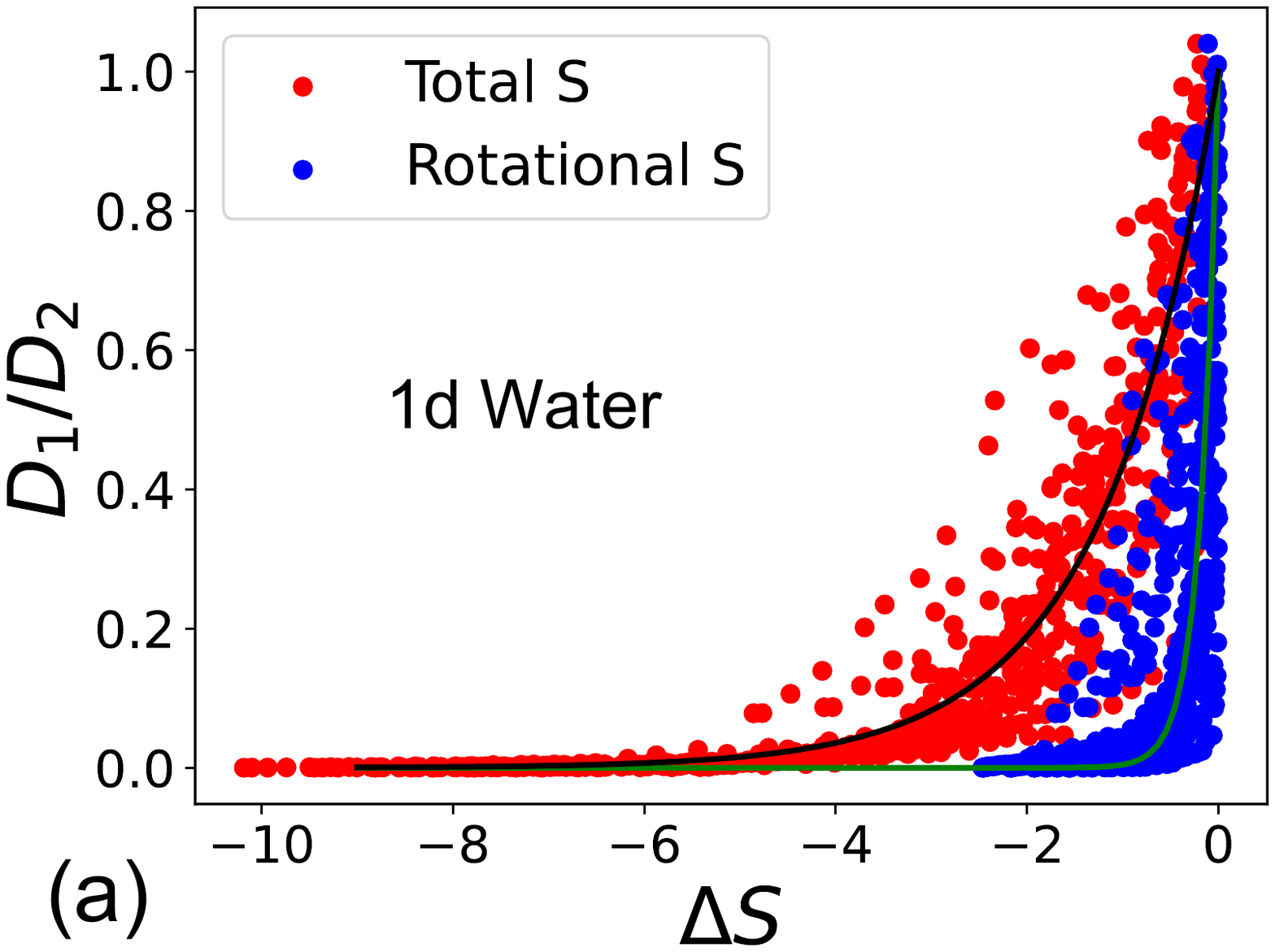}
  \includegraphics[width=0.7\textwidth, height=4cm]{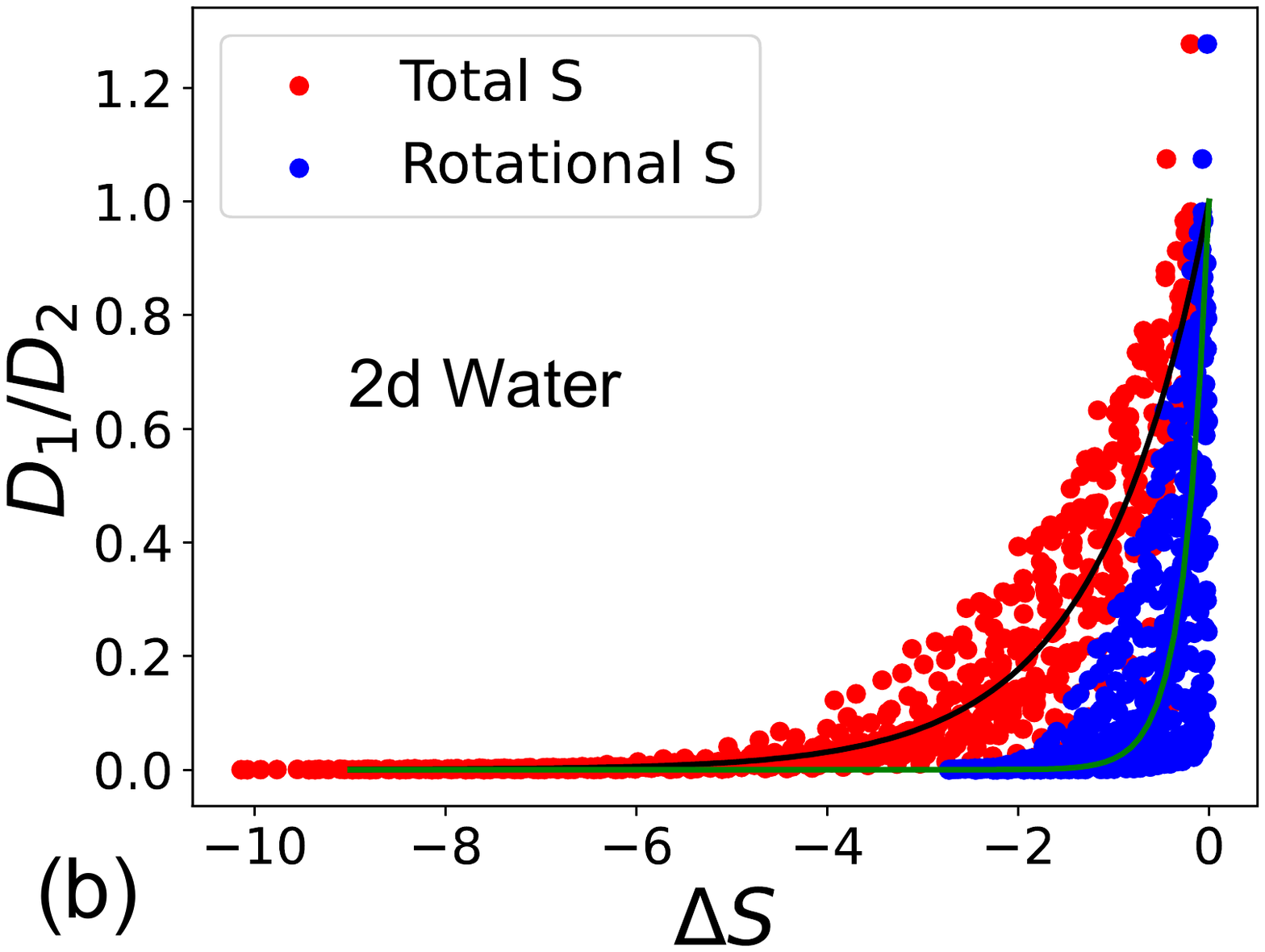}
  \includegraphics[width=0.7\textwidth, height=4cm]{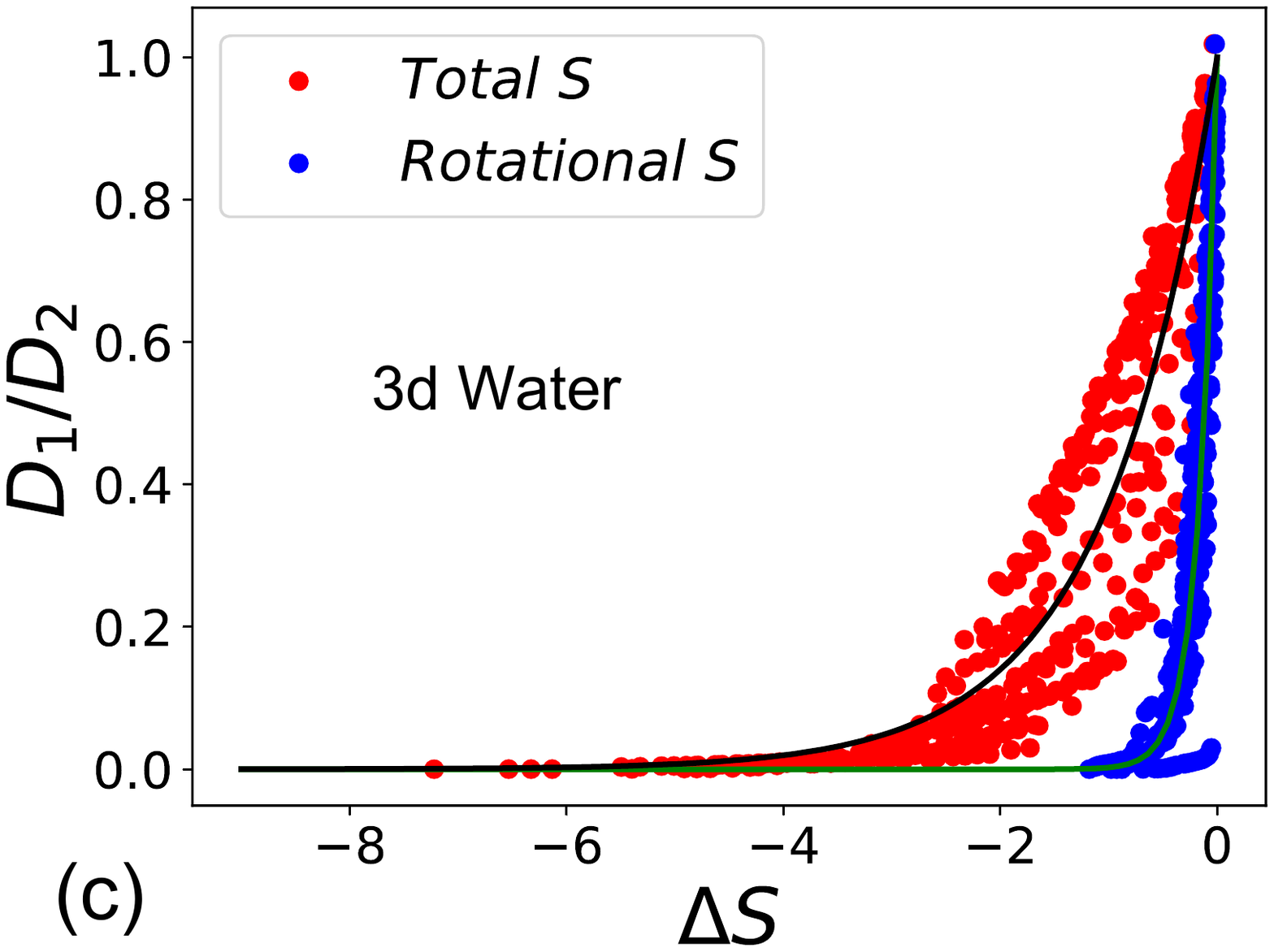}
  \label{fig:rot}
\end{minipage}%

\caption{The figures (a),(b), and (c) illustrate the two-state diffusion entropy relation for \(1d\), \(2d\), and \(3d\) liquid water, respectively. The red scatter plot illustrates the ratio of diffusion coefficients \(D_1/D_2\) between states 1 and 2 as a function of their total entropy difference \(\Delta S\). The blue scatter plot shows \(D_1/D_2\) between states 1 and 2 as a function of their rotational entropy difference. The black and green lines represent the data fitting curve of the form \(D_1/D_2=\exp[(\alpha/d) \Delta S]\) for total and rotational entropy difference, respectively. For the case of total entropy difference (black curve), the value of prefactor \(\alpha/d\) is 0.83, 0.86, and 0.98 for 1d, 2d, and 3d systems, respectively. For the case of rotational entropy difference (green curve), the value of prefactor \(\alpha/d\) is 5.399, 3.956, and 5.997 for \(1d\), \(2d\), and \(3d\) systems, respectively. The spread of data points is notably smaller for rotational entropy, with a similar trend observed for translational entropy, compared to the scatter plot of total entropy. This suggests that individual degrees of freedom, when considered separately, exhibit greater conformity to the diffusion entropy relation than total entropy.}
 \label{fig:rot}
\end{figure}
\clearpage
\bibliography{aipsamp}

\end{document}